\documentclass[12pt]{article} 

\makeindex
\begin{document}

\title{QUANTUM DEFORMATIONS OF RELATIVISTIC SYMMETRIES: \\
 SOME RECENT DEVELOPMENTS}

\author{J.LUKIERSKI
                   \\ 
Institute for Theoretical Physics, University
 of Wroc{\l}aw, \\pl. Maxa Borna 9, 50-204 Wroc{\l}aw, Poland\\
 E-mail: lukier@ift.uni.wroc.pl}

\date{}
\maketitle

\begin{abstract}
Firstly we discuss different versions of
noncommutative space-time and corresponding appearance of quantum
 space-time groups. Further we consider the relation between quantum
deformations of relativistic symmetries and so-called doubly
special relativity (DSR) theories.
\end{abstract}


\baselineskip=13.07pt
\section{Introduction}\label{irlus1}
Quantum deformations of Lie algebras and Lie group were motivated
by quantum universe scattering method and introduced in 1980`s as
noncocommutative Hopf algebras \cite{irlu1}$^{-}$\cite{irlu3}.
Subsequently the notion of quantum symmetries was tried for many
symmetries occurring in physics, in particular for the basic
relativistic symmetries, described by Poincar\'{e} algebra and
Poincar\'{e} group, as well as anti-de-Sitter (AdS), de-Sitter
(dS) and conformal symmetries.

Because the fourmomenta generators contrary to the Lorentz
rotations are dimensionfull, they introduce into the space-time
algebras the notion of scaling. One should distinguish two types
of quantum deformations:

i) The ones introducing \underline{dimensionless} deformation
parameter $q$, which is invariant under rescaling of the
fourmomenta. The prototypes of such deformations are provided by
Drinfeld-Jimbo (DJ) deformations of \cite{irlu1,irlu4} AdS, dS or
conformal algebra. One can show \cite{irlu5,irlu6} that it does
not exist DJ quantum deformation of Poincar\'{e} algebra,
obtained by the extension of $DJ$ deformation for the Lorentz
subalgebra.

The dimensionless deformation parameter of space-time symmetries
appears less attractive from the point of view of physical
applications. It is agreed that the quantum symmetries and
noncommutative space-time coordinates should become relevant for
very small distances (e.g. at Planck length $l_{p} \simeq
10^{-33}$cm).
 We conclude that the dimensional parameter in
quantum algebra structure will characterize the distances at which
the notions of classical geometry are not valid. Therefore we
should introduce.

ii) Second type of deformations of space-time symmetries with
built-in elementary length, or elementary mass. First such a
deformation has been proposed in 1991 \cite{irlu7}, with the
deformation parameter $\kappa$ (the classical limits is provided
by limit $\kappa \to \infty$), and as called the
$\kappa$-deformation. Still it is not clear how to relate by
rigorous proof the parameter $\kappa$ with the Planck mass $M_P$
($M_P \simeq 10^{19}$GeV), but due to the quantum gravity origin
of noncommutativity of space-time coordinates it is believed that
they are linked very closely, and quite often are assumed
 to be identical.

\section{Noncommutativity of Space-Time and Quantum
Groups}\label{irlus2}

The need of quantum space-time symmetries with dimensionfull
deformation parameter can be seen clearly from the
noncommutativity of space-time coordinates. The general relations
can be written as follows (see e.g. \cite{irlu8})
\begin{eqnarray}
[\widehat{x}_{\mu}, \widehat{x}_{\nu}]& =& \frac{1}{\kappa^{2}}
F(\kappa x)
= \frac{1}{\kappa^{2}} \Theta^{(0)}_{\mu \nu} \cr &&
 + \frac{1}{\kappa}
\Theta^{(1)}_{\mu \nu}{}^{\nu} \widehat{x}_{\nu} +
 \Theta^{(2)}_{\mu \nu}{}^{\rho \theta} \widehat{x}_{\rho}
 \widehat{x}_{\tau} + \ldots \, ,
 \label{irlu1}
\end{eqnarray}
where we introduced the parameter $\kappa$ in order to express
 the noncommutativity in terms of dimensionless
 coordinates $y_k = \kappa x_\mu$.

Let us consider the special cases when only one constant tensor
 $\Theta^{(k)}_{\mu\nu}{}^{\rho_{1} \ldots \rho_{k}}\neq 0$.

 1) $\Theta^{(0)}_{\mu\nu} \neq 0$,  $\quad
\Theta^{(k)}_{\mu\nu}{}^{\rho_{1} \ldots \rho_{k}}=0$, $\ k=1,2,3
\ldots$

This example was studied extensively recently; the Poincar\'{e}
symmetries
 are broken by a constant tensor but remain classical. Such a form
 of deformed space-time was obtained by Seiberg and Witten
 \cite{irlu9} by considering the space-time manifold as described
 by the $D$-brane world volume in the presence of constant tensor
 field $B{\mu\nu}$ (we recall that such a field is necessary for
 consistency of supergravity framework in $D$=10).

 2) $\Theta^{(1)}_{\mu\nu}{}^{\rho}\neq 0$,
 $\quad \Theta^{(k)}_{\mu\nu}{}^{\rho_{1}\ldots \rho_{k}}=0$,
 $ \ k=0,2,3,4 \ldots$.

 In such a case we obtain the Lie-algebraic form of deformed
 space-time algebra. It appears that the $\kappa$-Minkowski
 space-time, obtained in the framework of standard
 $\kappa$-deformations of Poincar\'{e} symmetries
 \cite{irlu10}$^-$\cite{irlu12} belongs to such a class of new theories. In
 1996 there was introduced the generalized $\kappa$-deformation of
 Poincar\'{e} symmetries\cite{irlu13}, with the choice
 \begin{equation}\label{irlu2}
 \Theta^{(1)}_{\mu\nu}{}^{\rho} =
 \frac{i}{\kappa}\left(
a_{\mu}\delta_{\nu}^{\ \rho} - a_{\nu}\delta_{\mu}^{\ \rho}
 \right)
 \, ,
\end{equation}
where $a_{\mu}$ denotes constant fourvector.

{}From (\ref{irlu2}) follows that the noncommuting quantum
direction in Minkowski space is described by the coordinate
$\widehat{y}=a^{\mu}x_{\mu}$. It has been also shown that the
classical $r$-matrix corresponding to (\ref{irlu2}) satisfies

- modified Yang-Baxter (YB) equation if $a^{2}_{\mu}\neq 0$

- classical YB equation if $a^{2}_{\mu}= 0$.

The quantum deformations of relativistic symmetries with quantized
light-cone direction ($a^{2}_{\mu}=0$) was firstly described by
Ballesteros et all \cite{irlu14} and called null-plane quantum
Poincar\'{e} symmetries. Further it has been shown \cite{irlu15}
that in such a case the quantization can be described by the
twisting procedure \cite{irlu16,irlu17}.

If $\Theta^{(1)}_{\mu\nu}{}^{\rho}\neq 0$, the translations
$\widehat{v}_{\mu}$
\begin{equation}\label{irlu3}
\widehat{x}_{\mu} \longrightarrow\widehat{x}_{\mu}^{\prime} =
\widehat{x}_{\mu} + \widehat{v}_{\kappa}\, ,
\end{equation}
described by the coproduct of $\widehat{x}_{\mu}$, are also
noncommutative
\begin{equation}\label{irlu4}
[\widehat{v}_{\mu}, \widehat{v}_{\nu}] = \Theta^{(1)}_{\mu\nu}
{}^{\rho}\widehat{v}_{\rho}\, .
\end{equation}

The extension of noncommutative translations (\ref{irlu4}) to
quantum Poincar\'{e} group is only possible for particular choices
of $\Theta^{(1)}_{\mu\nu}{}^{\rho}$, in particular for the one
given by (\ref{irlu2}). The general classification of quantum
Poincar\'{e} groups has been considered by Podle\'{s} and
Woronowicz \cite{irlu18}.

3) $\Theta^{(2)}_{\mu\nu}{}^{\rho\tau}\neq 0$, $\quad
\Theta^{(k)}_{\mu\nu}{}^{\rho_{1} \ldots \rho_{k}}=0$,
 $ \ k=0,1,3,4 \ldots$.

 In such a case the relation (\ref{irlu1}) does not contain any
 dimensionfull parameter and it describes the quantum deformation
 of relativistic symmetries with dimensionless deformation
 parameter (e.g if the Lorentz sector is described by
 Drinfeld-Jimbo deformation). Such deformed space-time framework
 is described by \underline{braided} quantum symmetries, because
 the quantum translations (\ref{irlu3}) do satisfy the relations
\begin{equation}\label{irlu5}
[\widehat{v}_{\mu}, \widehat{v}_{\nu}]=
\Theta^{(2)}_{\mu\nu}{}^{\rho\tau}\widehat{v}_{\rho}\widehat{v}_{\tau}\,
,
\end{equation}
but also one has to assume that
\begin{equation}\label{irlu6}
[\widehat{x}_{\mu},\widehat{x}_{\nu}]\neq 0\, .
\end{equation}

The first example of braided quantum Poincar\'{e} group has been
presented by Majid \cite{irlu5}.

\section{Nonlinear Realizations of Relativistic Symmetries Versus Quantum
 Deformations}\label{irlus3}

 There are two sources of the modification of relativistic
 symmetries (see e.g.  \cite{irlu19}).

 i) One can change nonlinearly the basis of classical Poincar\'{e}
 algebra
 \begin{eqnarray}\label{irlu7}
[M^{(0)}_{\mu\nu}, M^{(0)}_{\rho\tau}] & = &
 i(\eta_{\mu\rho}\, M^{(0)}_{\nu\tau} + \ldots \, ,
 \cr
[M^{(0)}_{\mu\nu}, P^{(0)}_{\rho}] & = &
 i(\eta_{\mu\rho}\, P^{(0)}_{\nu}
 - \eta_{\nu\rho} P^{(0)}_{\mu} \, ,
\cr [P^{(0)}_{\mu}, P^{(0)}_{\nu}] & = & 0 \, ,
  \end{eqnarray}
  by introducing the \underline{deformation map} - the invertible
  nonlinear functions of the generator. An important special class
   of deformation maps described only the change of four momentum
   basis
   \begin{equation}\label{irlu8}
P^{(0)}_{\mu}\longrightarrow P_{\mu} = P_{\mu}(P^{(0)};\kappa) =
F_{\mu}^{\ \nu} \left( \frac{P^{(0)}}{\kappa}\right)
P^{(0)}_{\nu} \, ,
 \end{equation}
 while the Lorentz generators remain unchanged
 ($M_{\mu\nu} =M^{(0)}_{\mu\nu}$).
 Introducing the \underline{inverse deformation map}

\begin{equation}\label{irlu9}
P^{(0)}_{\mu}= P_{\mu}^{(0)} (P;\kappa)= \widetilde{F}_{\mu}^{\
\nu} \left( \frac{P}{\kappa} \right) P_{\nu}\, ,
\end{equation}
we see that the \underline{mass Casimir} is modified as follows:
\begin{equation}\label{irlu10}
C_{1} = P^{(0)}_{\mu} \, P^{(0)\mu}= P_{\mu}^{(0)}  \left(
\frac{P}{\kappa} \right) P_{\mu}^{(0)\mu}  \left(
\frac{P}{\kappa} \right) = m^{2}_{0} \, ,
\end{equation}
i.e. we obtained \underline{deformed nonlinear} energy-momentum
dispersion
 relation.
 Other consequence of the deformation map is the nonlinear modification
 of energy-momentum addition and conservation laws.
  The primitive coproduct for the generators $P^{(0)}_{\mu}$ is replaced
  by
\begin{equation}\label{irlu11}
\Delta(P_{\mu}) = P_{\mu}\left( P_{\mu}^{(0)}\otimes 1 + 1
\otimes  P_{\mu}^{(0)}, \kappa \right)\,.
\end{equation}
Denoting for two-particle system
\begin{description}
\item{$P_{\mu}(i)$}  -   the fourmomenta of i-th particle ($i=1,2$)

\item{$P_{\mu}(1,2)$}  -   the fourmomenta of 2-particle system
\end{description}
and using (\ref{irlu10}), (\ref{irlu8}) one can express the
coproduct (\ref{irlu11}) as describing nonlinear energy-momentum
addition law \cite{irlu20,irlu21}
\begin{equation}\label{irlu12}
P_{\mu}(1,2) =P_{\mu} \left( P^{0)}_{\mu} (P(1);\kappa) +
P^{0)}_{\mu}(P(1);\kappa);\kappa \right)
\end{equation}

The composition law (\ref{irlu12}) is symmetric, what indicates
 that we are dealing with classical Lorentz symmetries
 nonlinearly realized in the four-momentum sector.

 The choice of the deformation map which provides the deformed
 mass Casimir in bicrossproduct basis of $\kappa$-deformed
 Poincar\'{e} algebra \cite{irlu11,irlu12}

\begin{equation}\label{irlu13}
C_{1} = \left( 2\kappa \sinh \frac{P_0}{2\kappa} \right)^{2}
-e^{-\frac{P_0}{\kappa}}\overrightarrow{P}^{2} = M^{2}\, ,
\end{equation}
leads to Doubly Special Relativity theory (DSR) of
Amelino-Camelia et all
 (see e.g.  \cite{irlu22}).
Such a choice of deformed mass Casimir implies maximal value of
three-momentum if energy $\frac{E}{c}=P_0 \to \infty$.

Indeed, in simple case $M^2 = 0$ one gets
\begin{equation}\label{irlu14}
\overrightarrow{P}^{2}= \kappa^2 \left( 1 - e^{-
\frac{P_0}{\kappa}} \right)^{2}
\mathop{\longrightarrow}\limits_{P_{0} \to \infty }   \ \kappa^2
\end{equation}
i.e. we obtain the operational definition of mass-like deformation
parameter $\kappa$ as maximal possible value of the
three-momentum. It appears that selecting properly the
deformation map
 (\ref{irlu8}) and corresponding nonlinear mass Casimirs (\ref{irlu10})
  one can introduce three different variants of DSR theories \cite{irlu23}.

ii)  One can modify relativistic symmetries by introducing quantum
 deformations of Hopf algebra describing Poincar\'{e} symmetries.

  The quantum deformations of relativistic symmetries can be easily
   distinguished from the classical relativistic symmetries in nonlinear
    disguise if we observe that

    -- quantum deformations imply \underbar{nontrivial}
    \underbar{ bialgebra} \underbar{structure},
     described by classical $r$-matrix ($I_i\equiv (M^{(0)}_{\mu\nu},
     P^{(0)}_{\mu}$); $I_i \wedge I_j \equiv I_i \otimes I_j - I_j \otimes
     I_i$)
     \begin{equation}\label{irlu15}
r= a^{ij}\, I^{(0)}_{i}\wedge I^{(0)}_{j}\,.
     \end{equation}
     The calssical $r$-matrix describes infinitesimal deformation of
     classical coproduct $\Delta^{(0)}$
\begin{equation}\label{irlu16}
\Delta(I^{(0)}_{i}) = \Delta^{(0)}(I^{(0)}_{i}) +
\xi[r,\Delta^{(0)}(I^{(0)}_{i})]+ {\cal O}(\xi^{2}) \, .
\end{equation}
The formula (\ref{irlu16}) implies that the coproduct is not
symmetric.

-- If the relativistic symmetries are quantum-deformed, the
space-time coordinates are not commuting, contrary to the case in
DSR framework.

We see therefore that there are simple criteria to distinguish
between the
 modification due to change of classical basis and the one following from
 genuine quantum deformations [ , ].

 The formula (\ref{irlu16}) can be "integrated" to arbitrary vales
 of deformation parameter $\xi$ if we introduce the similarity
 transformation
\begin{equation}\label{irlu17}
\Delta_{F}(I^{(0)}_{i}) = F \circ \Delta^{(0)} (I^{(0)}_{i})
\circ F^{-1}\, ,
\end{equation}
where $F=F^{(1)}_{i} \otimes F^{(2)}_{i}$ is the \underline{twist
function}
 with the following linear term in the power expansion in $\xi$
\begin{equation}\label{irlu18}
F = 1 \otimes 1 + \xi \, a^{ij} \, I_i \otimes I_j + {\cal
O}(\xi^{2})\, .
\end{equation}
The twist function $F$ leads to cassociative coproducts
(\ref{irlu17})
 if it satisfied the equation
\begin{equation}\label{irlu19}
F_{12} \left( \Delta^{(0)}\otimes 1 \right) F = F_{23} \left( 1
\otimes \Delta^{(0)} \right) F \, .
\end{equation}
Expanding (\ref{irlu19}) in powers of $\xi$ one gets from
  the bilinear terms the classical Yang-Baxter equation. Further
  twist quantization modifies the converse (antipode) as follows
\begin{equation}\label{irlu20}
S_{F} (I^{(0)}_{i}) = u S^{(0)} (I^{(0)}_{i}) u^{-1}\, ,
\end{equation}
where $u=F^{(1)}_i \cdot S F^{(2)}_i$.

The twist quantization changes only the coproducts and coinverse
- the
   classical Lie algebra relations and the counit remain
 unmodified.

 \section{Quantum Deformations of AdS and Conformal Algebras}
 Drinfeld twist quantization method can be applied to any deformation described
 infinitesimally  by classical $r$-matrix satisfying CYBE.
  Recently there were  explicitly written down the classical $r$-matrices for
  $O(3,2)$ and $O(4,2)$ algebras with generators belonging to the Borel
  subalgebra $B_{+}$ and subsequently these bialgebras were
  quantized \cite{irlu24,irlu15}.

  Let us consider as an example $O(3,2)$ algebra. If we introduce the
  Cartan-Weyl basis for $O(3,2)\simeq Sp(4)$ (see e.g. \cite{irlu7})
\begin{equation}\label{irlu21}
\mathop{\underbrace{h_{1}, h_{2}}}\limits_{Cartan\atop
generators}, \mathop{\underbrace{e_{\pm 1}, e_{\pm
2}}}\limits_{simple\, root \atop generators},
\mathop{\underbrace{e_{\pm 3}, e_{\pm 4}}}\limits_{composite \
root \atop generators}
\end{equation}
the most general classical $O(3,2)$ $r$-matrix support in $B_+
\otimes B_+$  is
 the following:
 \begin{equation}\label{irlu22}
r=\alpha [(2h_1 + h_2) \wedge e_4 + 2e_1 \wedge e_2 ] + \xi h_2
\wedge e_2 + \rho e_2 \wedge e_4 \,.
 \end{equation}
The corresponding twist function has been calculated in
\cite{irlu24} and is the
 product of four factors: Jordanian twist, extended Jordanian twist, deformed
  Jordanian twist and Reshetikhin twist.

  It should be pointed out that the generators of $O(3,2)$ can be
  physically assigned to D=3 conformal or D=4 AdS algebra.
   In first conformal case one
  can show that the parameters $\alpha, \xi$ and $\rho$ are
   dimensionfull, and  we arrive at
    the $\kappa$-deformation of D=3 conformal algebra
     \cite{irlu24}. The assignment of D=4
    AdS generators leads to different conclusion: the deformation parameters
 can remain dimensionless because the role of dimensionfull parameter is taken over
      by the AdS radius.

      The twist quantization of $O(4,2)$ algebra
 interpreted physically         as D=4 conformal algebra has been
      considered in \cite{irlu15}, where all classical $r$-matrices
       with support in Borel subalgebra were quantized.
 In particular in \cite{irlu15} the light-cone $\kappa$-deformation of
  Poincar\'{e} algebra (see (\ref{irlu2}) with $a^2_{\mu}=0$) has
   been extended to the particular $\kappa$-deformation of D=4
    conformal algebra.
       The alternative physical interpretation of twisted $O(4,2)$ as
 quantum D=5 AdS algebra can be found in  \cite{irlu25}.

      \section{Final Remarks}
      The formalism of twist quantization has been recently extended to
       all classical superalgebras\cite{irlu26}, in particular to
      orthosymplectic superalgebras $OSp(n;2m)$. For $n=1,m=2$ one obtains
      in such a way new deformations of D=4 AdS superalgebra. Subsequently,
       using suitable contraction method, one can obtain twist quantization
        of D=4 Poincar\'{e} superalgebra.



\begin{thebibliography}{99}
\bibitem{irlu1} V.G. Drinfeld, in {\it Quantum Groups}, Proc. Int.
Congress of Mathematics, Berkeley, USA, 1986, p. 798.

\bibitem{irlu2} S.L. Woronowicz, Comm. Math. Phys. {\bf 111}, 613 (1987).

\bibitem{irlu3} L.D. Faddeev, N.Yu. Reshetikhin and
 L.A. Takhtayan, Leningrad Math. Journ. {\bf 1}, 193 (1990).

\bibitem{irlu4} L.Faddeev, N. Resetikhin and L. Takhtajan,
 Alg. Anal. {\bf 1}, 178 (1990).

\bibitem{irlu5} S. Majid, J. Math. Phys. {\bf 34}, 2045 (1993).


\bibitem{irlu6} P. Podle\'{s}, S.L. Woronowicz, in: "Proceedings of First
 Caribbean School on Mathematics and Theoretical Physics", Guadeloupe 1993, ed.
 R. Coquereaux, World Scientific (1995), p. 364.

\bibitem{irlu7} J. Lukierski, A. Nowicki, H. Ruegg and V.N. Tolstoy,
 Phys. Lett. {\bf B264}, 331 (1991).
\bibitem{irlu8} P. Kosi\'{n}ski, J. Lukierski, P. Ma\'{s}lanka,
 Czech. J. Phys. {\bf 50}, 1283 (2000); Phys. Atom Nucl. {\bf 64}, 2139 (2001).

\bibitem{irlu9} N. Seiberg, E. Witten,

\bibitem{irlu10} S. Zakrzewski, J. Phys. {\bf A27}, 2075 (1994).

\bibitem{irlu11} S. Majid, H. Ruegg, Phys. Lett. {\bf B334}, 348 (1994).

\bibitem{irlu12}J. Lukierski, H. Ruegg, W.J. Zakrzewski,  Ann. Phys.
{\bf 243}, 90 (1995), (hep-th/9312153 ).

\bibitem{irlu13}  P. Kosi\'{n}ski and P. Ma\'{s}lanka, in ``From Quamtum
Field Theory to Quantum Groups", ed. B. Jancewicz, J. Sobczyk,
World Scientific, 1996, p. 41; q-alg/9512018.


\bibitem{irlu14}A. Balesteros, F.J. Herranz, M.A. del Olmo and M.
Santander, Phys. Lett. {\bf B351}, 137 (1995).


\bibitem{irlu15} V. Lyakhowski, J. Lukierski, M. Mozrzymas, Phys.
Lett. {\bf B538}, 375 (2002).

\bibitem{irlu16} V. Drinfeld, Dokl. Acad. Nauk SSSR, {\bf 273}, 531 (1983).

\bibitem{irlu17} P.P. Kulish, V.D.
and A.I. Mudrov, J. Math. Phys. {\bf 40}, 4569 (1999).

\bibitem{irlu18} P. Podle\'{s}, S.L. Woronowicz, Commun. Math. Phys. {\bf
178}, 61 (1996), (hep-th/9412059).

\bibitem{irlu19} J. Lukierski, A. Nowicki, Czech. J. Phys.
 {\bf 52}, 1261 (2002); hep-th/0209017.

\bibitem{irlu20} J. Lukierski, A. Nowicki,
Int. J. Mod. Phys. {\bf 18}, 7 (2003), hep-th/0203065.

\bibitem{irlu21}J. Judes and M. Visser, qr-qc/0205067.

\bibitem{irlu22}  G. Amelino-Camelia, Phys. Lett. {\bf B510}, 255 (2001);
  Int. J. Mod. Phys. {\bf D11}, 35 (2002).


\bibitem{irlu23} J. Lukierski, A. Nowicki, hep-th/0210111.

\bibitem{irlu24} V. Lyakhovski, J. Lukierski, M. Mozrzymas, Mod. Phys.
 Lett. {\bf A18}, 753 (2003).

\bibitem{irlu25} A. Ballesteros, N.R. Bruno, F.J. Herranz, Phys.
Lett. {\bf B574}, 273 (2003).
\bibitem{irlu26} V. Tolstoy, math.QA/0402433.


\end{thebibliography}
\end{document}